\documentclass[10pt, conference, letterpaper]{IEEEtran}
\ifCLASSINFOpdf
\else
\fi
\usepackage{amsmath}
\usepackage{amsfonts}
\usepackage{xcolor}
\usepackage{tikz}
\usetikzlibrary{arrows}
\usepackage{mathtools}
\usepackage{algorithm}
\usepackage{algorithmic}

\makeatletter
\newcommand\fs@norules{\def\@fs@cfont{\bfseries}\let\@fs@capt\floatc@ruled
  \def\@fs@pre{}%
  \def\@fs@post{}%
  \def\@fs@mid{\kern3pt}%
  \let\@fs@iftopcapt\iftrue}
\makeatother
\floatstyle{norules}
\restylefloat{algorithm}

\makeatletter
\newcommand{\mytag}[2]{%
  \text{#1}%
  \@bsphack
  \protected@write\@auxout{}%
         {\string\newlabel{#2}{{#1}{\thepage}}}%
  \@esphack
}
\makeatother

\DeclarePairedDelimiter\floor{\lfloor}{\rfloor}

\begin{document}
\title{A note on how the problem of Partion of Integers show in Caching}

\author{\IEEEauthorblockN{Mohit Thakur}
\IEEEauthorblockA{Institute of Communications and Navigation\\
DLR (German Aerospace Center)\\
Oberpfaffenhofen-Wessling, Germany\\
Email: mohit.thakur@dlr.de}
%\and
%\IEEEauthorblockN{Homer Simpson}
%\IEEEauthorblockA{Twentieth Century Fox\\
%\and
%\IEEEauthorblockN{James Kirk\\ and Montgomery Scott}
%\IEEEauthorblockA{Starfleet Academy\\
%San Francisco, California 96678--2391\\
%Fax: (888) 555--1212}
}


%Telephone: (800) 555--1212, Fax: (888) 555--1212}


\maketitle

\begin{abstract}
In this article, we show how the finding the number of partitions of same size of a positive integer show up in caching networks. We present a stochastic model for caching where user requests (represented with positive integers) are a random process with uniform distribution and the sum of user requests plays an important role to tell us about the nature of the caching process. We discuss Euler's generating function to compute the number of partitions of a positive integer of same size. Also, we derive a simple approximation for guessing the guessing the number of partitions of same size and discuss some special sequences. Lastly, we present a simple algorithm to enumerate all the partitions of a positive integer of same size.
\end{abstract}


% \fi
%
% For peer review papers, this IEEEtran command inserts a page break and
\IEEEpeerreviewmaketitle

\section{Introduction}\label{intro}

Caching in networks is considered as an alternative to convert device memory into data rate to reduce delay and improve quality-of-service (QoS). The topic has been studied for a particular information theoretic setting in \cite{maddah-ali-fund-limits-caching} and for decentralized setting in \cite{maddah-ali-decentralized-caching}. Since then, a lot of work has been done in coded-storage, device-to-device etc. with respect to caching. A good account of it can be found in \cite{caire-molisch-D2D} and \cite{caire-debbah-businesscase-caching}.

\begin{figure}\label{fig:cachemodel}
\centering
    \begin{tikzpicture}
        \draw [black] (-1,0) circle [radius=0.3]; \node at (-1,0) {UE};
        \draw [black] (3,0) circle [radius=0.3]; \node at (3,0) {S};
        \draw [black] (-0.7,0) to (2.7,0);
        \draw [->,>=stealth] (-1.3,-1) -- (-1.3,-0.5); \node at (-1,-1.3) {requests};
        \draw [->,>=stealth] (-1,-1) -- (-1.0,-0.5); %\node at (-1.0,-1.3) {\footnotesize{$F_5$}};
        \draw [->,>=stealth] (-0.7,-1) -- (-0.7,-0.5); %\node at (-0.6,-1.3) {\footnotesize{$F_{11}$}};
        \node at (-0.3,-0.8) {$\cdots$};
        \draw [black] (2.5,-0.5) to (2.5,-1.5) to (3.5,-1.5) to (3.5,-0.5);
        \draw [black] (2.5,-0.7) to (3.5,-0.7);
        \draw [black] (2.5,-0.9) to (3.5,-0.9);
        \draw [black] (2.5,-1.1) to (3.5,-1.1);
        \draw [black] (2.5,-1.3) to (3.5,-1.3);
    \end{tikzpicture}
    \caption{Cache model.}
\end{figure}
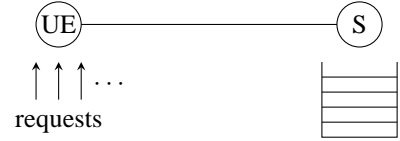

In this article, we consider a user that uses a device to access content from the internet. The device is also referred to as a user equipment (UE). Suppose the device is connected to a server that can store (or cache) a limited amount of content in its memory. Such a server can be located within the UE or at a distant location as a separate entity. Figure~\ref{fig:cachemodel} shows such a system with user requests arriving at UE that are partially or wholly served by the server S. Caching has many advantages such as improved quality of service (QoS). An important facet to study in this situation is how can we understand the nature of user requests for the content in relation to the cached content.

The problem can be modeled in a straightforward way using a stochastic setting. Let $F=\{F_1, F_2, \ldots, F_N\}$ be a set of files with and $[K]=\{1, \ldots,K\}$ be the set of first $K$ positive integers. At a given instant the user requests a file from $F$ at random. Suppose the stochastic process of user requests is given by $X=(X_{1}, X_{2},X_{3}, \ldots)$, where $X_{k} \in F$ and the probability that $X_k$ is some file $F_j \in F$ is given by $P(X_{k}=F_j)=p_j$. An associated stochastic process $I=(5,1,7,21, \ldots)$ of file indices can be easily constructed from the user request process $X$, where the k$^{\text{th}}$ element takes the value $j \in [N]$ if $X_k=F_j$. Now, if we ask the following questions of the following type
\begin{enumerate}
    \item what is the probability that a finite sub-sequence $\widehat{I} \subset I$ sums to a positive integer $M$?
    \item does a finite sub-sequence exits such that the sum of it's elements equal's $S$?
\end{enumerate}

What the above questions have in common is that they require us to know the number of all possible partitions of integer $S= I_n+ \cdots +I_{n+k}$ into $k$ parts such that each part belongs to $[N]$. In the next section we describe some classical results from the theory of partition of integers that gives a generating function for the problem.

\section{Generating function for Integer Partitioning}\label{cachnintpart}
Let $S$ be a positive integer that is a sum of $k$ parts where each part belongs to $[N]$, i.e.,
\begin{align*}
    S= r_1 + \cdots +r_k,
\end{align*}
where $r_j \in [N]$ for $1 \leq j \leq k$.

The coefficient of $x^S$ in the Euler's generating function
\begin{align*}
    \frac{1}{(1+x+x^2+\cdots)(1+x^2+x^4+\cdots)(1+x^3+x^6+\cdots) \cdots}
\end{align*}
gives us the number of partition of $S$ for all $k$. The general form of the Euler's generating function is given by
\begin{align*}
    \mathcal{G}(x,y)=\frac{1}{(1-yx)(1-yx^2)(1-yx^3) \cdots}.
\end{align*}
Here, the coefficient of $x^Sy^k$ is the total number of partitions of $S$ in $k$ parts. See sectio $3.16$ in \cite{genfunc-wilf} for a detailed discussion.

Now we can play with some special cases that interests us. Consider two sets, $U$ consisting of non-negative integers and $V$ of positive integers. For instance, let $U=\{0,1,2,..,u\}$ and $V=[N]$. Then the generating function of the partitions of $S$ in $k$ parts belonging to the set $V$ and allowed multiplicities from $U$ is given by
\begin{align*}
    \mathcal{G}(x,y,U,V) = \displaystyle\prod_{v \in V} \left( \displaystyle\sum_{u \in U} y^{u}x^{uv} \right).
\end{align*}
Even though $G(x,y,U,V)$ is a polynomial with finite terms, it is in general a NP-Complete problem to compute the coefficients of powers of $xy$. Next we present some special cases, when it is easier to approximate the coefficient of $xy$.

\subsection{Sums with large $k$ and large $N$}
Consider a stochastic process $I$ as described in section~\ref{intro} and that each element is chosen uniformly from the set of positive integers $\mathbf{Z}^{+}$. Let $S_{n-m}$ denote the sum of elements
\begin{align*}
    S_{n-m} = I_{m+1}+\cdots+I_{n}.
\end{align*}
The mean and variance of $I_k$ are given by
\begin{align*}
    \mu(I_k) & = \frac{1}{N}\displaystyle\sum_{i=1}^{N} i = \frac{N+1}{2} \\
    \sigma^2(I_k) & = \frac{1}{N}\displaystyle\sum_{i=1}^{N} i^2 - \mu^2(I_k)= \frac{(8N+6)(N^2 - 1)}{24}
\end{align*}
Now, the central limit theorem tells that for any real $\alpha < \beta$
\begin{align*}
    \begin{split}
        P \left( \alpha  < \frac{S_{n-m} - (n-m)\mu(I_k)}{\sigma(I_k)\sqrt(n-m)} < \beta\right) & \\
         \longrightarrow \mathfrak{N}(\beta)  - \mathfrak{N}(\alpha), &
    \end{split}
\end{align*}
as $n-m$ goes to infinity.
Therefore, we have
\begin{align}
    \begin{split}
        & P \left( (\alpha \sigma(I_k)\sqrt{n-m}+(n-m)\mu(I_k)) \right.  \\
        & \left.  < S_{n-m} < (\beta \sigma(I_k)\sqrt{n-m}+(n-m)\mu(I_k)) \right)  \\
        & \longrightarrow \mathfrak{N}(\beta)  - \mathfrak{N}(\alpha)  \label{berdiffCLT1}
    \end{split}
\end{align}
as $n-m$ goes to infinity. Let $\mathcal{N}(S,k,U,[N])$ denote the number of partitions of $S$ in $k$ parts, where each part belongs to $[N]$ with multiplicities from $U$. Then from inequality~\ref{berdiffCLT1} we get
\begin{align}\label{cltapprox}
    \mathcal{N}(S,k,U,[N]) \approx \frac{(n-m)!}{\floor*{\frac{n-m}{2}}!} \frac{1}{P(\alpha,\beta,S_{n-m})},
\end{align}
where $P(\alpha,\beta,S_{n-m})$ denotes the probability
\begin{align*}
    \begin{split}
        & P \left( (\alpha \sigma(I_k)\sqrt{n-m}+(n-m)\mu(I_k)) \right.  \\
        & \left.  < S_{n-m} < (\beta \sigma(I_k)\sqrt{n-m}+(n-m)\mu(I_k)) \right).
    \end{split}
\end{align*}

\emph{Remark}: The approximation~\ref{cltapprox} calls for the following remarks.
\begin{enumerate}
    \item Approximation~\ref{cltapprox} is bounded by the error term that only vanishes as $(n-m)$ tends to infinity.
    \item The first term of right hand side of approximation~\ref{cltapprox} accounts for that for large sums with large $(n-m)$, on the average $(n-m)/2$ elements tends to be the same. Hence, the term accounts for the total number of permutations for a partition.
\end{enumerate}

\subsection{Special Sums}
Let the sub-sequence of elements be of length $m$ with $k$ elements of value $N-1$ and $m-k$ elements with value $N$, where $0 \leq k\leq m$. Then the sum
\begin{align*}
    S = k(N-1)+(m-k)N=mN-k
\end{align*}
cannot be partitioned in $m$ parts using any other element from $[N]$, other than $N-1$ and $N$. Therefore, there exist only one partition $S$ with $m$ parts.

Furthermore, as elements of the sub-sequence take the values uniformly and randomly from $[N]$, the sub-sequence can occur in
\begin{align*}
    \frac{m!}{k!}
\end{align*}
ways. Therefore, the probability of a sub-sequence that is $m$ elements long and sums to $S$ is
\begin{align*}
    \frac{k!}{m!}.
\end{align*}

\section{Enumeration of Partitions of Same Size}
In this section we present a simple algorithm to enumerate the partitions a positive integer into $m$ parts.

Consider again a positive integer $S$ that is to be divided in $m$ parts, where each part belongs to the set $[N]$.  And let $\mathcal{N}(S,k,[N])$ denote the total number of such partitions. Then we have
\begin{align*}
    S = k.d+r.
\end{align*}
Here, $0 \leq r <d$ and $d \in [N]$. This means the $S$ can be written as a sub-sequence
\begin{align}\label{enualgseq1}
    (\underbrace{d,\ldots,d}_{\mytag{$\hat{k}$}{termA}},\underbrace{d+1,\ldots,d+1}_{\mytag{$\hat{r}=m-\hat{k}$}{termA}}),
\end{align}
where all the elements sum up to $S$. We can rewrite the sub-sequence as follows
\begin{align}\label{enualgseq2}
    (\underbrace{d_1, \ldots, d_1}_{\mytag{$k$}{termC}}, d_2,\underbrace{d_3,\ldots, d_3}_{\mytag{$r$}{termD}}),
\end{align}
where $d_1$, $d_2$ and $d_3$ are chosen according to
\begin{enumerate}
    \item Decrease the left most element that is greater than $1$ and increase the right-most element that is less than $N$, until the sequence becomes either $(1,\ldots,d)$, $(1, \ldots, 1, d, N, \ldots, N)$ or $(N-1, \ldots, N-1, N,\ldots, N)$, where $1 \leq d \leq N$. Of course, for the third case there is only sequence that partitions $S$.
\end{enumerate}

From the sequence~\ref{enualgseq2}, another sequence of the same sum can be constructed by simply by reducing right-most $r$ elements by $1$ and distributing the value $r$ to the rest of the elements by increasing the elements up to $d_3-1$ starting from the k$^{th}$, then $k-1^{th}$ element and so on. For instance, the sequence $(1,1,1,5,8,8)$ has $k=3$ and $r=2$, and after the process it becomes $(1,1,1,7,7,7)$ while the sum of two sequences remain the same. Denote the values of the $i^{th}$ element sequence by $e_i$ and let  
\begin{align*}
    \hat{r} & =\displaystyle\sum_{j=1}^{m-r} (e_{m}-1 - e_j) \\
    \text{ and } \hat{m} & = \begin{cases}
                                m \text{ if $\hat{r}-r \geq 0$}, \\
                                m-(\hat{r}-r) \text{ otherwise.}
                                \end{cases}
\end{align*}
If $\hat{r}-r = -r' < 0$, then the process can be applied to the first $\hat{m} = m-r'$ elements of the sequence while keeping the last $r'$ elements the same. Otherwise, $\hat{m}$ equals $m$. Let us call this process as the \emph{enumeration process}. The following algorithm describes the enumeration process for a given $\hat{m}$.

\begin{algorithm}[H]
 \caption{Enumeration Process}
 \begin{algorithmic}[1]
 \renewcommand{\algorithmicrequire}{\textbf{Input:}}
 \renewcommand{\algorithmicensure}{\textbf{Output:}}
 \REQUIRE $\mathcal{S}=\{(d_1, \ldots, d_1, d_2,d_3,\ldots, d_3)\}$, $r$ and $\hat{m}$
 \ENSURE  $\mathcal{N}(S,m,[N])$
 %\\ \textit{Initialisation} :
  %\\ \textit{LOOP Process}
  %\PROCEDURE
  \STATE Evaluate $\hat{r}=\displaystyle\sum_{j=1}^{m-r} (e_{m}-1 - e_j)$
  \REPEAT
  \IF {$\hat{r} -r \geq 0$}
        \FOR {$l=1$ to $\hat{m}-r-1$}
            \STATE $e_{m-r-1}=min(e_{m-r-1}+r,e_m -1)$
            \STATE $e_{m-r-l}=min(1+[r-l(e_m-1)-e_{m-r-1}]_{\mathbf{I}},e_m -1)$
            \STATE $e_{m-r+1}= \cdots=e_{m-r+\hat{m}}=e_{m}-1$
            \STATE $\mathcal{N}(S,m,[N])= \mathcal{N}(S,m,[N])+1$
            \STATE $r \gets min \left(\hat{m},r+\floor*{\frac{r}{d_3-1-d_2}}+ \floor*{\frac{r-(d_3-1-d_2)}{d_3-1}}\right)$
            \STATE $\hat{m} \gets \displaystyle\sum_{i=1}^{m}(e_m - e_i)$
            \RETURN $s_{i}=(e_1, \ldots, e_m)$
            \STATE $\mathcal{S} \gets \mathcal{S} \cup s_{i}$
        \ENDFOR
        \ELSE
            \STATE $m \gets m- (\hat{r} -r-1)$
            \STATE go to $1$
        \ENDIF
        \UNTIL {$e_{r} - e_{r-1}=1$}
        \RETURN $\mathcal{S}$
 \end{algorithmic}
 \end{algorithm}

If we repeat this process for $\hat{m} \in [m]$, we get the algorithm that enumerates all partitions of $S$ that have $m$ parts each.

\begin{algorithm}[H]
 \caption{Enumeration algorithm for integer partitions of same size}
 \begin{algorithmic}[1]
 \renewcommand{\algorithmicrequire}{\textbf{Input:}}
 \renewcommand{\algorithmicensure}{\textbf{Output:}}
 \REQUIRE $\mathcal{S}^{m}$ and $m$
 \ENSURE  $\mathcal{N}(S,m,[N])$
 %\\ \textit{Initialisation} :
  %\\ \textit{LOOP Process}
  
  \FOR {$\hat{m}=m$ to $\hat{m}=2$}
    \STATE Enumeration Process$(\mathcal{S})$     
  \ENDFOR
 \RETURN $\mathcal{N}(S,m,[N])=|\mathcal{S}^{m}|$
 \end{algorithmic}
 \end{algorithm}

It is straightforward to see that the algorithm $2$ is exhaustive and enumerates all the partitions of $S$ of size $m$ while counting the number of partitions. We leave out the proof out for brevity. In general, the enumeration algorithms do not serve much purpose in determining a useful formula for computing the number of partitions except for the special case when the problem instances are really small, but they can be a source of insight to the nature of the problem.

%\section*{Acknowledgment}



\bibliographystyle{IEEEtran}
\bibliography{./../Bib/IEEEabrv,./../Bib/Bib_Probability,./../Bib/Bib_Combinatorics,./../Bib/Bib_Caching}

\begin{thebibliography}{1}
\providecommand{\url}[1]{#1}
\csname url@samestyle\endcsname
\providecommand{\newblock}{\relax}
\providecommand{\bibinfo}[2]{#2}
\providecommand{\BIBentrySTDinterwordspacing}{\spaceskip=0pt\relax}
\providecommand{\BIBentryALTinterwordstretchfactor}{4}
\providecommand{\BIBentryALTinterwordspacing}{\spaceskip=\fontdimen2\font plus
\BIBentryALTinterwordstretchfactor\fontdimen3\font minus
  \fontdimen4\font\relax}
\providecommand{\BIBforeignlanguage}[2]{{%
\expandafter\ifx\csname l@#1\endcsname\relax
\typeout{** WARNING: IEEEtran.bst: No hyphenation pattern has been}%
\typeout{** loaded for the language `#1'. Using the pattern for}%
\typeout{** the default language instead.}%
\else
\language=\csname l@#1\endcsname
\fi
#2}}
\providecommand{\BIBdecl}{\relax}
\BIBdecl

\bibitem{maddah-ali-fund-limits-caching}
M.~A. Maddah-Ali and U.~Niesen, ``Fundamental limits of caching,'' \emph{{IEEE}
  Trans. Inf. Theory}, vol.~60, pp. 2856--2867, 2014.

\bibitem{maddah-ali-decentralized-caching}
------, ``Decentralized coded caching attains order-optimal memory-rate
  tradeoff,'' \emph{{IEEE/ACM} Trans. Netw.}, vol.~23, pp. 1029--1040, 2015.

\bibitem{caire-molisch-D2D}
G.~C. Mingyue~Ji and A.~F. Molisch, ``Fundamental limits of caching in wireless
  d2d networks,'' \emph{{IEEE} Trans. Inf. Theory}, vol.~62, pp. 849--869,
  2016.

\bibitem{caire-debbah-businesscase-caching}
E.~B. Georgios~Paschos, I.~Land, G.~Caire, and M.~Debbah, ``Wireless caching:
  Technical misconceptions and business barriers,'' \emph{arXiv}, 2016.

\bibitem{genfunc-wilf}
H.~S. Wilf, \emph{Generatingfunctionology}.\hskip 1em plus 0.5em minus
  0.4em\relax San Diego: Academic Press, Inc., 1994.

\end{thebibliography}
\end{document}